# Engineering the thermal conductivity along an individual silicon nanowire by selective helium ion irradiation

Yunshan Zhao[1,†], Dan Liu[1,2,†], Jie Chen[3], Liyan Zhu[4], Alex Belianinov[5], Olga S. Ovchinnikova[5], Raymond R. Unocic[5], Matthew J. Burch[5], Songkil Kim[5], Hanfang Hao[1,2], Daniel S Pickard[1,2], Baowen Li[6,*], John T L Thong[1,2,*]

[1] Department of Electrical and Computer Engineering, National University of Singapore, Singapore 117583, Republic of Singapore

[2] NUS Graduate School for Integrative Sciences and Engineering, National University of Singapore, Singapore 117456, Republic of Singapore

[3] Center for Phononics and Thermal Energy Science, School of Physics Science and Engineering, Tongji University, Shanghai 200092, People's Republic of China

[4] School of Physics and Electronic & Electrical Engineering, Huaiyin Normal University, Jiangsu 223300, People's Republic of China

[5] Center for Nanophase Materials Sciences, Oak Ridge National Laboratory, Oak Ridge, Tennessee 37831, USA.

[6] Department of Mechanical Engineering, University of Colorado, Boulder 80309, USA

[†] These authors contributed equally to this work.

**Abstract**

The ability to engineer the thermal conductivity of materials allows us to control the flow of heat and derive novel functionalities such as thermal rectification, thermal switching, and thermal cloaking. While this could be achieved by making use of composites and metamaterials at bulk scales, engineering the thermal conductivity at micro- and nano-scale dimensions is considerably more challenging. In this work we show that the local thermal conductivity along a single Si nanowire can be tuned to a desired value (between crystalline and amorphous limits) with high spatial resolution through selective helium ion irradiation with a well-controlled dose. The underlying mechanism is understood through molecular dynamics simulations and quantitative phonon-defect scattering rate analysis, where the behavior of thermal conductivity with dose is attributed to the accumulation and agglomeration of scattering centers at lower doses. Beyond a threshold dose, a crystalline-amorphous transition was observed.

The thermal conductivity of bulk materials can be engineered by various means such as the use of composite materials[1] or metamaterials[2]. This allows the manipulation of heat flow with precision and inspires various functionalities, like thermal inverter, thermal concentrator, thermal cloak, *etc*[3-11]. However, the realization of such concepts at the nanoscale is difficult because engineering the thermal conductivity of materials at small dimensions is much more difficult. The thermal conductivity of nano-sized material systems can be tuned by phonon-boundary scattering[12,13], interfacial thermal resistance[14], and/or by exploiting phononic crystals[15,16], but to engineer the thermal conductivity locally with precision is challenging from both fabrication and theoretical-prediction aspects. In particular, in the case of Si nanowires – a versatile building block for nanoelectronic[17], photovoltaic[18], thermoelectric[19] and photonic[20] devices – it has been shown experimentally that the thermal conductivity of an individual Si nanowire can be reduced by shrinking its diameter[13,21] and by introducing surface roughness[19,22,23]. However, these prior works only demonstrated the tuning and measurement of the thermal conductance of a nanowire *as a whole* between its ends. Significantly, in terms of mechanical robustness for *practical* device design, ultra-thin or highly-roughened silicon nanowires are extremely fragile, apart from the fact that making reliable ohmic contacts to such silicon nanowires is also a major issue[24]. Moreover, the use of physical structuring means that post-fabrication tuning of thermal conductance by this method is not possible.

Another approach to engineer the thermal conductivity of Si nanowire without affecting its surface morphology is through ion implantation. By introducing impurity atoms into the crystalline silicon nanostructures, the thermal conductivity can be tuned due to enhanced phonon - defect scattering, and the damage can be annealed away (at CMOS process-compatible thermal budgets) as well[25]. Even though heavy-ion irradiation (e.g. using gallium and uranium ions) has been observed to change the thermal conductivity in nanostructures by defect engineering[26-28], point-defect generation is inevitably accompanied by cluster amorphization along the ion trajectory. For less heavy ions, like silicon ions, the process of continuously elastic collision would make it difficult to generate a spatially uniform defect profile, and a complicated region containing simple defects as well substantial cluster formation results[25,29]. It is thus desirable to adopt a technique that can produce different concentrations of simple point defects in a well-controlled manner in order to establish unequivocally the role of such defects on the thermal conductivity.

In this paper we show that the thermal conductivity of an individual Si nanowire can be locally changed by irradiating it with helium ions with well-defined doses at different positions along its length (see Methods for detailed sample preparation). Because of the small nanowire diameter (~160 nm) and moderately high helium ion energy (30-36 keV), the helium ions pass through the nanowire with minimal forward scattering and energy loss, and their effect is reasonably uniformly distributed over the nanowire cross-section, unlike the case of a much thicker Si substrate. The thermal conductivity of each irradiated nanowire segment was then measured by an electron beam heating technique that is capable of spatially resolving the thermal conductivity along the nanowire's length[30] (Methods present the detailed measurement and calculation procedure). Using this technique, the effect of different helium ion dose can be measured conveniently along a single nanowire, thereby eliminating potential discrepancies that could arise from sample-to-sample variations in the nanowire diameter and from uncertainty in the thermal contact resistance between the nanowire ends and the thermometers. From the measurement we observed a reproducible thermal conductivity versus ion irradiation dose curve, from which a clear crystalline-to-amorphous transition is seen. The effect of defects created by helium ions below the transition dose threshold on the nanowire thermal conductivity is discussed with theoretical support from the kinetic theory of phonon gas model and non-equilibrium molecular dynamics (NEMD) simulations. We also show that the thermal conductivity of the irradiated nanowire can be further tuned by annealing. The ability to tune the thermal conductivity along silicon nanowires with high spatial resolution by a well-controlled helium ion irradiation, and to measure the local thermal conductivity thereafter, provides a new platform to study nanoscale thermal transport.

**Results**

**Thermal conductivity measurement of individual Si nanowire irradiated by HIM.** The silicon nanowire is suspended between two temperature sensors comprising platinum (Pt) loops on silicon nitride membranes, each of which is suspended by 6 nitride beams. Figure 2(a) shows the measurement result for Sample #1 for the damaged portion irradiated with the highest dose ($7.5\times10^{16}$ cm$^{-2}$). From this figure we can see that as the electron beam scans from the left to right across the damaged portion, the temperature rise of the left (right) sensor, $\Delta T_L$ ($\Delta T_R$), undergoes obvious decrease (increase), indicating that the thermal resistivity of the damaged portion is much larger than that of the intrinsic portion. This is further confirmed by the increase in slope of the $R_i(x)$ curve within the damaged portion, where $R_i(x)$ is the cumulative thermal resistance from the left sensor to the heating spot. It is observed that the power absorption of the nanowire from the electron beam, $(\Delta T_L+\Delta T_R)/R_b$, where $R_b$ is the equivalent thermal resistance of the suspension beams, does not vary significantly at the damaged portion, indicating negligible material removal from inside the nanowire. The $R_i(x)$ curves for all the portions with different doses are plotted in Supplementary Fig. 3. The thermal conductivity of each portion was calculated as $\kappa = \frac{1}{\frac{dR_i}{dx}\cdot A}$, where $A$ is the cross-sectional area of silicon nanowire with diameter $d=160$ nm as measured in the TEM. To minimize the error, $\frac{dR_i}{dx}$ was obtained by linearly fitting the $R_i(x)$ curve of the damaged portion and the intrinsic portion, with the beginning and ending 50 nm of these portions stripped off in order to avoid the non-uniformity in the dose near the boundaries arising from ion forward scattering. As the thermal conductivity is derived from the gradient ($\frac{dR_i}{dx}$) of each portion, one can conveniently obtain the local thermal conductivity of each irradiated/non-irradiated portion in a single nanowire. Moreover, the need to contend with the unmeasurable thermal contact resistance between the ends of the nanowire and the sensors, which is a long standing problem of the conventional thermal bridge method, is avoided with our technique.

Plotted in Figure 2(b) are the thermal conductivity versus dose curves for the eight samples measured at room temperature. The error bar is calculated by error propagation due to the uncertainty in $\frac{dR_i}{dx}$ during linear fitting and the uncertainty in the diameter (±5 nm), with the former being one to two orders smaller than the latter. The intrinsic thermal conductivity of the nanowire is 50.2±2.3 Wm$^{-1}$K$^{-1}$, obtained by linearly fitting several undamaged portions of the measured samples and taking the average. The amorphous limit of ~1.7 Wm$^{-1}$K$^{-1}$ (at 300 K) is taken from the literature[31,32].

Clearly two regimes can be identified in Figure 2(b) – in the regime of relatively low dose ($<1.5\times10^{16}$ cm$^{-2}$), the thermal conductivity drops rapidly initially and then tapers off as the dose increases, whereas in the regime of dose $\geq 3\times10^{16}$ cm$^{-2}$, the thermal conductivity approaches an asymptotic value. Of particular note is a steep jump in thermal conductivity between these two regimes, as shown in the inset of Figure 2(b).

**Low dose regime – point defect creation and agglomeration.** The as-implanted damage was calculated by Monte Carlo simulations based on a binary collision approach (TRIM/SRIM[33], the detailed simulation conditions and results are described in Supplementary Fig. 2 and Supplementary Note 1). The simulation shows that one helium ion can create a large number of displaced atoms (or Frenkel pairs, each of which is composed of a vacancy and a nearby interstitial) along its track; moreover, this as-produced damage is discrete and distributes relatively uniformly across the nanowire cross-section due the large range of the light and high energy (30-36 keV) incoming helium ions compared to the sample thickness ($d$=160 nm); lastly, from simulation results, 98% of the helium ions penetrate the sample, and one helium ion can create 33 defects on average, while the number of residual (embedded) helium atoms is much smaller than the number of damaged lattice sites.

However, at room temperature, the as-produced point defects are not stable[34]. They undergo extensive interstitial-vacancy recombination, defect clustering, and pairing with impurities (e.g., carbon and oxygen) during and after the irradiation process at room temperature. As a result, only 1%-10% of the defects survive at room temperature for bulk materials[35]. For the case of nanowires, annihilation at the surface may also enhance this annealing effect given the high surface-to-volume ratio[36].

An upper bound for the concentration of defects can be roughly estimated by assuming that 10% of the Si vacancies created by helium ions survive. For calculation purposes, we take it that the Si nanowire is circular with a diameter of 160 nm, based on our experimentally imaged and measured cross-section. The number of vacancies remaining is a function of dose and this relation is finally described as $2.0\times10^5$ (cm$^{-1}$) × dose (cm$^{-2}$), with details discussed in detail in Supplementary Note 1.

As there are $5\times10^{22}$ Si atoms per cm$^3$, for a dose of $1\times10^{16}$ cm$^{-2}$, around 4% of the Si atoms are displaced, acting as scattering centers for phonons. This value provides an upper bound of the phonon scattering centers because we use the upper bound of the survival rate (10%). Moreover, there is a consensus view that the stable defects left in the room-temperature silicon are predominantly divacancies[37], which are formed by two nearby vacancies. It has also been shown experimentally that at room temperature, divacancy is the dominant defect for the near-surface region of the helium-ion damaged Si[38]. Regarding its impact on the thermal conductivity, this means *two* displaced sites combine and act as *one* phonon scattering center. Besides the vacancy-type defect, there is also a small amount of helium ions left; however, the quantity is limited as most of the helium ions penetrate the nanowire during irradiation. There may also be interstitial or interstitial-like defects in the nanowire, because although Si interstitials can follow athermal migration even at 4.2 K and the migration distance is long in high purity material, they can also form clusters or complexes which are stable at room temperature[34].

The inverse phonon life time (phonon relaxation rate) contributed by *substitutional* point defects for a simple cubic lattice is $\tau_d^{-1} = D\omega^4$, with

$$D = c \frac{a^3}{4\pi v^3}\left(\frac{\Delta M}{M}\right)^2 \quad (1)$$

where $c$ is the point defect concentration per atom, $\Delta M$ the mass difference between the substitutional atom and the host atom, $M$ the mass of host atom, $a$ the lattice constant and $v$ the phonon group velocity[39,40]. For the case of vacancy defect, as it removes the mass of one atom and also the force constants of two atomic sites, it can be treated as[41,42] $\frac{\Delta M}{M} = -3$. For simplicity and better comparison with previous studies, we use Equation 1 for the substitutional point defect calculation in silicon nanowires, since this formula captures the essence of phonon scattering rates due to vacancy defects, and it is indeed valid for complex lattices[43,44]. Hence, compared with the substitutional point defects, the vacancy defect has a much stronger phonon-scattering effect and is highly effective in reducing the lattice thermal conductivity. Indeed, experiments by Pei *et al.*[45] showed that a small amount of vacancy defects (3%) in the $In_2Te_3$ lattice introduced by alloying with InSb can substantially reduce the lattice thermal conductivity (by ~80%).

In order to investigate the effect of point defects on the phonon scattering rate, we first consider a simple Debye formula for the thermal conductivity: $\kappa = \frac{1}{3}Cv^2\tau$, where $C$ is the specific heat, $v$ is the average group velocity assuming a linearized phonon dispersion model, and $\tau$ is the phonon relaxation time. We further assume that $\tau^{-1}$ can be calculated from Matthiessen's rule, which implies that different phonon scattering mechanisms are independent of each other. As a result, the thermal resistivity ($1/\kappa$) is additive from various scattering mechanisms, including $\tau_d^{-1}$, the phonon-defect scattering rate. Thus $1/\kappa$ should be linearly related to the *concentration of point defects*, as $\tau_d^{-1}$ depends linearly on the concentration of point defects (Equation 1). However in the experiment what can be obtained is the $1/\kappa$ curve as a function of *dose* (Figure 3(a)), from which we can see that in the low dose regime, $1/\kappa$ firstly increases linearly and then tends towards saturation.

It is possible to circumvent the linearized-phonon-dispersion assumption and extract the value of parameter $D$ by fitting the measured thermal conductivities within the framework of the kinetic theory of phonon gas[46]. The fitting model, utilizing the *full phonon dispersion* of Si, includes anharmonic phonon-phonon scattering events and the scattering of phonon by isotopes, boundaries and point defects as well (a detailed description can be found in Supplementary Note 2). The optimal $D$ obtained from the best fit is plotted as a function of dose as shown in Figure 3(b), where the possible influence from boundary scattering is considered separately. The exclusion of boundary scattering term makes the parameter $D$ increase by a factor of 4 compared with that of the case where boundary scattering is included. Nevertheless the overall trend for $D$ with dose is qualitatively similar -- the parameter $D$ shows a linear dependence on dose in the low-dose regime and becomes saturated in the high-dose regime. The relationship between $D$ and dose can be well fitted by a simple function, $D = E(1 - e^{-F \cdot dose})$. Where the dose is small, the parameter $D$ scales linearly with dose, namely, $D = E \cdot F \cdot dose$. Considering the fact that the phonon-defect scattering rate depends linearly on the concentration of point defects (Equation 1), we may infer that for doses less than $1 \times 10^{15}$ cm$^{-2}$ in our experiment, the point defects created by helium ion irradiation is linearly proportional to the helium ion dose. However, as the dose increases the parameter $D$ converges gradually to some asymptotical value, which suggests that the effective number of scattering centers does not increase significantly with further increases in dose. One possible reason is that for higher dose (from $2 \times 10^{15}$ to $1 \times 10^{16}$ cm$^{-2}$), the point defects agglomerate due to room temperature annealing. Lastly, the parameter $D$ is about 1-3 orders of magnitude larger than the corresponding parameter for isotope scattering[46], which indicates the dominant role of point defects in suppressing the thermal conductivity of Si nanowire.

To further examine the role of point defect scattering, and to incorporate the effect of phonon boundary scattering, NEMD simulation was carried out for qualitative comparisons with the experimental results. In the simulation, Si nanowires with various cross-sections of 3×3, 6×6 and 9×9 unit cells (0.543 nm per unit cell) and fixed length of 20 unit cells were considered (Supplementary Fig. 6). Point defects are in the form of vacancies modeled by randomly removing some Si atoms (up to 20%) and then removing the single-bonded atom pairs (Supplementary Fig. 7). The details of the NEMD simulation are given in Supplementary Note 3.

Compared to the pristine Si nanowires, the simulated thermal conductivity of the defective Si nanowires is reduced, and decreases with the increase of vacancy concentration due to the enhanced phonon scattering by vacancy (Figure 3(c)). At low vacancy concentration (smaller than 5%), the thermal conductivity of defective Si is further reduced by boundary scattering, while it is almost the same for different diameters at high vacancy concentration (greater than 15%), indicating the dominant role of vacancy scattering in this regime. We also plotted the $1/\kappa$ curve as a function of vacancy concentration (Figure 3(d)). As the Si nanowire diameter used in the simulation increases, $1/\kappa$ shows no trend of saturation as the *isolated* vacancy concentration increases; on the other hand this corroborates the model of point defect agglomeration in the real situation under helium ion irradiation at room temperature.

From the experimental and simulation results we can see that in the low dose regime, point defects are effective in changing the thermal conductivity of the helium-irradiated Si nanowire. Firstly, the thermal conductivity initially decreases drastically as the dose (and the corresponding density of point defects) increases; however further increasing the dose yields diminishing return on reducing thermal conductivity, i.e., in order to obtain the same marginal decrease of thermal conductivity, one has to at least double the dose (inset of Figure 2(b)). Secondly, at most 4% displaced Si atoms (corresponding to a dose of $1\times10^{16}$ cm$^{-2}$) is capable of decreasing the nanowire thermal conductivity by nearly 70%. It should be noted that under our experimental conditions of high helium ion energy, thin sample, and same dose rate and temperature, the thermal conductivity for a particular dose is repeatable for all the samples measured.

**Effect of annealing.** It has been shown in the literature that point defects that are stable at room temperature such as divacancies can be annealed out at higher temperatures[37,47]. This annealing effect on the thermal conductivity was investigated using Sample #3. Upon annealing, the total thermal resistance ($R_T$) of the nanowire became progressively smaller. The results from spatially resolved thermal conductance measurement show that this decrease of $R_T$ is due to the increase of the irradiated-portion's thermal conductivity, while the thermal conductivity of the non-irradiated portion remained the same (Figure 4(a)) as expected.

The before- and after- annealing results are plotted in Figure 4(b). From this figure we can see that for annealing at 120 °C, the thermal conductivity of the irradiated portion increases noticeably after 20 hours. To further improve the nanowire thermal conductivity at 120 °C, a long annealing time, e.g., 160 hours, is required. On the other hand, merely 2 hours of annealing at 300 °C improved the thermal conductivity substantially, and for the lowest dose ($2.5 \times 10^{14}$ cm$^{-2}$), the thermal conductivity had almost recovered to that of the non-irradiated portions. This is further illustrated by Figure 4(c), in which the $R_i(x)$ curve becomes a straight line, and the irradiated and non-irradiated potions are indistinguishable from each other.

It has been shown that the predominant point defects – divacancies – become mobile only at 200-300 °C, but at lower temperatures, divacancies could be annihilated by interstitials that have dissociated from clusters and diffused through the lattice. It was previously found that by annealing a proton-irradiated Si substrate at 150 °C for ~60 hours, the divacancy concentration can be reduced by 54%[48]. This annihilation of divacancies by interstitials explains the increase in the thermal conductivity upon annealing at 120 °C. On the other hand, at 300 °C, divacancies become mobile and migrate to sinks (primarily to the surface in the case of a nanowire), and if their concentration is high enough, they also agglomerate to form small clusters that are stable up to 500 °C[37]. This explains the experimental results of the 300 °C annealing – at low doses ($2.5\times10^{14}$ cm$^{-2}$), nearly all the divacancies migrate to sinks and disappear, so that the thermal conductivity has practically fully recovered, while at higher doses, divacancies agglomerate, forming clusters which continue to act as scattering centers, so that the thermal conductivity of these portions could not be recovered. Figure 4(b) shows that the thermal conductivity of the irradiated Si nanowire can be further tuned through annealing.

**High dose regime - crystalline-amorphous transition.** There is a steep drop in thermal conductivity when the dose further increases from $2.5\times10^{16}$ cm$^{-2}$ (inset of Figure 2(b)). This demarcation between the low and high dose regimes indicates a phase transition in the material. We believe this phase change is the crystalline-amorphous transition of the Si nanowire arising from helium ion irradiation. For defect clusters and amorphous region in crystalline silicon nanowires, the phonon would be scattered differently. The defect clusters, which can be assumed to possess inner boundary if their size is large enough, have a relatively weak scattering for the phonon transport and the projected areas of defect clusters towards the heat-flow direction usually overlap, leading to the relative insensitive dependence of phonon transport on cluster size[49]. Hence the thermal conductivity tapers off as the dose increases (for dose below the critical value of amorphous region creation). In an amorphous silicon region, the phonon would be scattered significantly by disorder. For amorphous silicon, the majority of heat carriers are diffusons, which are non-propagating and delocalized modes, and their contribution is predicted to be ~1 Wm$^{-1}$K$^{-1}$, according to the calculation by Allen-Feldman theory[50,51]. The rest of the contribution is from the propagating propagons, which account for the remaining half of the thermal conductivity, due to their small proportions (around 3%). Thus, the existence of an amorphous region in defect-clustered crystalline silicon, if any, would greatly impede phonon transport.

In the low-dose regime, the disorder created by helium ions is mainly point defects, which agglomerate as the dose increases and the continuous aggregation of defects would change the crystalline lattice structure gradually in the silicon nanowires. When the dose goes beyond a critical value, the high energy ion-induced disorder can build up in crystalline silicon until its local free energy overcomes the energy barrier between the crystalline and amorphous phases. At this point, the defective crystalline lattice can collapse into amorphous phases[34]. Once an amorphous region is formed, it promotes the growth of damage in surrounding regions which in turn leads to further amorphization[52]. Finally, the entire irradiated region is amorphized.

This amorphization process can be visualized from the diffraction pattern of an irradiated Si nanowire prepared separately on a TEM grid. Figure 5(a) shows the bright-field image and diffraction pattern of an individual Si nanowire irradiated at two doses ($7.5\times10^{16}$ cm$^{-2}$ and $2.5\times10^{16}$ cm$^{-2}$), using a focused electron beam with a spot size of 200 nm. The non-irradiated portion of the nanowire shows diffraction spots of single-crystal Si (inset(c)); for the portion with irradiation dose of $7.5\times10^{16}$ cm$^{-2}$, the nanowire is fully amorphous, leaving only short-range order among Si atoms, as indicated by the Debye-Scherrer rings[25] (inset(b)); for the irradiated portion with dose of $2.5\times10^{16}$ cm$^{-2}$, both Debye-Scherrer rings and diffraction spots can be seen (inset(d)), indicating the coexistence of both amorphous and crystalline phases. Similar diffraction patterns for Sample #2 can also be seen in Supplementary Fig. 4(a). The coexistence of both amorphous and crystalline phases explains why the thermal conductivity for doses at $1\times10^{16}$ cm$^{-2}$ is higher than that for larger doses -- at the dose of $2.5\times10^{16}$ cm$^{-2}$, both crystalline and amorphous phases co-exist. However, the amorphous region rapidly takes over the entire volume for higher doses, such as the doses of $3.5\times10^{16}$ cm$^{-2}$ and $4\times10^{16}$ cm$^{-2}$, as shown in Supplementary Fig. 4(b), where the Debye-Scherrer rings become more obvious. At the dose of $5\times10^{16}$ cm$^{-2}$, the thermal conductivity is close to the amorphous limit. Our experiment also demonstrates that helium ion irradiation can amorphize the Si nanowire without changing its morphology. Moreover, the interface between the irradiated area (for dose of $7.5\times10^{16}$ cm$^{-2}$) and non-irradiated area is particularly sharp and distinguishable, which corresponds well to the thermal resistance measurement by the e-beam technique. This clear boundary confirms the well-controlled dose irradiation with definition of a few nanometers. The interface thermal resistance of these abrupt boundaries can be measured, and can be exploited to further reduce the thermal conductivity of an intrinsic nanowire by incorporating multiple such boundaries, but this is beyond the scope of the present study. Fig. 5(e) shows the clear interface and insets (f) and (g) are the FFTs for the crystalline and amorphous regions, respectively.

## Discussion

In this paper we have demonstrated that the thermal conductivity of an individual Si nanowire can be changed by selective helium ion irradiation. A single Si nanowire was irradiated at different positions with well-controlled helium ion doses, and an electron beam heating technique was used to measure the local thermal conductivity along the nanowire, which was then related to helium ion dose.

We observed a clear transition from crystalline Si to amorphous phase at a dose between $1.5\times10^{16}$ and $2.5\times10^{16}$ cm$^{-2}$ from the thermal conductivity versus dose curve. This result suggests a novel method to amorphize a Si nanowire without affecting its morphology. Moreover, within the dose regime in which only point defects are created, the Si nanowire thermal conductivity decreases drastically as the dose increases, and merely ~4% defects could reduce the thermal conductivity by ~70%, indicating a strong phonon scattering effect by the point defects. Within this regime, the effective scattering centers, inferred from parameter $D$, initially increase linearly with dose, and then saturates for larger dose. Finally, we observed that annealing could improve the thermal conductivity of the damaged nanowire, and at 300 °C for two hours, the thermal conductivity of a damaged portion irradiated with a dose of $1\times10^{15}$ cm$^{-2}$ could be recovered to the value of the non-irradiated case.

## Methods

**Sample preparation.** Silicon nanowires with diameter ~160 nm (Sigma-Aldrich 730866) grown along the [111] direction dispersed in 2-propanol solution were drop-casted onto a SiO$_2$/Si substrate. A single Si nanowire was picked up from the substrate by a nano-manipulator (Kleindiek MM3A-EM) with a sharp tungsten probe and placed on a pre-fabricated suspended METS device. The METS device comprises two silicon nitride membranes, each of which is suspended via six long suspension beams for thermal isolation. Integrated on top of the silicon nitride platforms are Pt loops acting as resistance temperature sensors. The nanowire was positioned so as to bridge the two sensors, with the two ends fixed onto the two sensors by the electron beam-induced deposition (EBID) of Pt-C composite. The EBID process was carried out using a focused electron beam with acceleration voltage of 30 kV and current of 1.3-5 nA.

The on-device Si nanowire was then cleaned using an Evactron RF plasma cleaner attached to the SEM chamber operated at a power of 14 W in 0.4 Torr of air for 2 hours, after which it was put into a helium ion microscope (Zeiss Orion Plus) chamber and pumped overnight. It was then irradiated by helium ions (~30-36 keV, 0.2-0.7 pA) with different doses at different positions (Figure 1(a)). This was achieved with the help of a nanometer pattern generation system.

Eight samples were irradiated under similar helium ion energy and current (or dose rate) at room temperature. The SEM images of samples #2 to #8 are shown in Supplementary Fig. 1 and the irradiation details are shown in Supplementary Table 1. The highest dose ($7.5\times10^{16}$ cm$^{-2}$) chosen is such that there is no significant sputtering of Si atoms, and a change in the Si nanowire diameter is not observable in a transmission electron microscope (TEM). The dose was then reduced for each irradiated position. The irradiation length and irradiation distance are different for all the measured silicon nanowires, the details of which are summarized in Supplementary Table 1.

Sample #3 was annealed in air at 120 °C for cumulative periods of 20, 40, 80 and 160 hours, with measurement by an electron beam heating technique carried out for each interval. After this, it was put into a tube furnace and annealed in forming gas at 300 °C for 2 hours. To check for possible oxidation during annealing, HRTEM and diffraction pattern measurements of the post-annealed nanowire were carried out and the results are shown in Supplementary Fig. 5. The nanowire is observed to remain single-crystalline and an unavoidable intrinsic oxide layer of ~2.5nm thickness is seen, which is similar to or even slightly smaller than that found in electroless-etching (EE) silicon nanowires[19].

**Measuring the thermal conductivity of each portion using electron beam heating technique.** In the electron beam heating technique, a focused electron beam is used as a localized heating source. By moving the electron beam along the nanowire and measuring the temperature rise ($\Delta T_L$ and $\Delta T_R$) of the left and right sensors corresponding to the position of the electron beam $x$ as measured from the left sensor, the cumulative thermal resistance $R_i(x)$ between the electron beam location and the left sensor is[30]:

$$R_i(x) = R_b \left\{ \frac{\alpha_0 - \alpha_i(x)}{1 + \alpha_i(x)} \right\} \quad (2)$$

where $\alpha_i(x) = \Delta T_L(x)/\Delta T_R(x)$, and $R_b$ is the equivalent thermal resistance of the six beams that link each suspended membrane to the environment as measured by the conventional thermal bridge method in which a 10 µA DC current was passed through the left Pt loop, raising its temperature ($\Delta T_{L0}$) by ~8.5 K, and the temperature rise of both sensors ($\Delta T_{L0}$ and $\Delta T_{R0}$) were measured from the resistance change of the Pt loops. Finally, $\alpha_0 = \Delta T_{L0}/\Delta T_{R0}$.

From the $R_i(x)$ curve, the spatially-resolved thermal conductivity of the nanowire can be calculated as: $\kappa = \frac{1}{\frac{dR_i(x)}{dx} \cdot A}$, where $A = \pi d^2/4$ and $d$ is the diameter of the nanowire.

## Data availability

The data that support the findings of this study are available from the corresponding authors upon request.

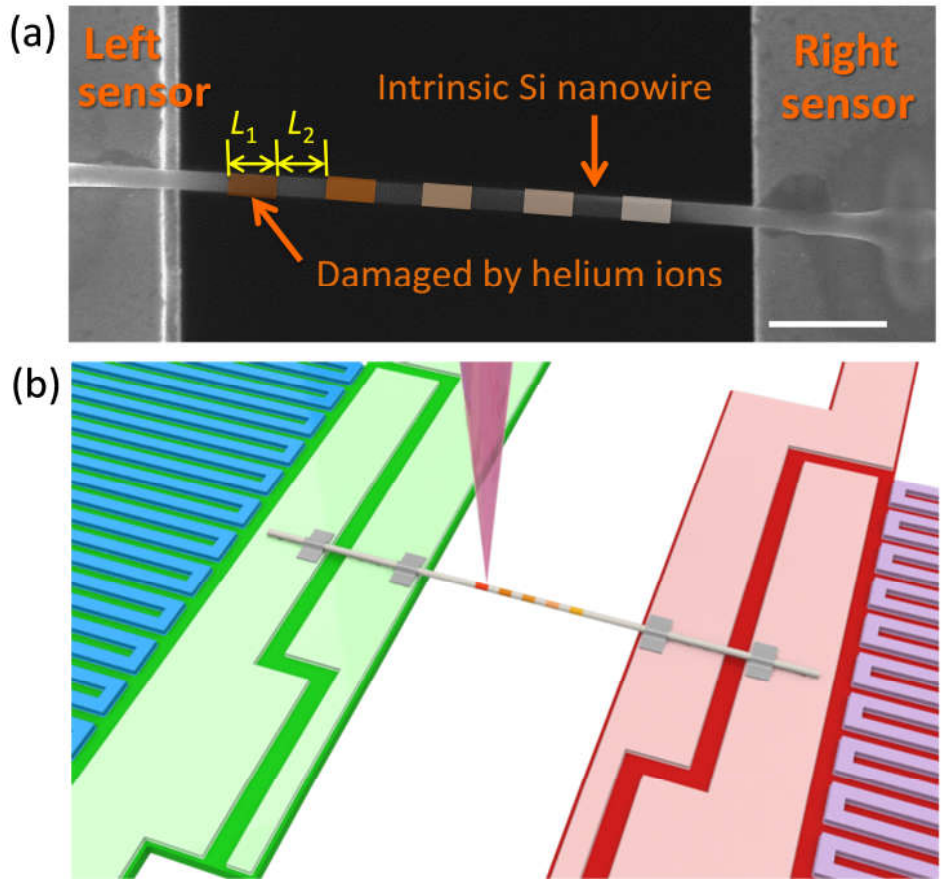

**Figure 1 | Silicon nanowire on METS device with helium ion irradiation. (a)** Scanning electron microscope (SEM) image of Si nanowire damaged by helium ions (Sample #1). The portions colored orange (with length of $L_1$) denote the parts damaged by helium ions; the uncolored portions (with length of $L_2$) denote the intrinsic Si nanowire. Scale bar is 1 μm. **(b)** Schematic showing the spatially resolved measurement of thermal conductivity of the damaged Si nanowire using electron beam heating technique. A focused electron beam (purple cone) is used as a heating source, and the temperature rise of the left and right sensors ($\Delta T_L$ and $\Delta T_R$) is measured by the Pt resistance thermometers (blue and purple loops).

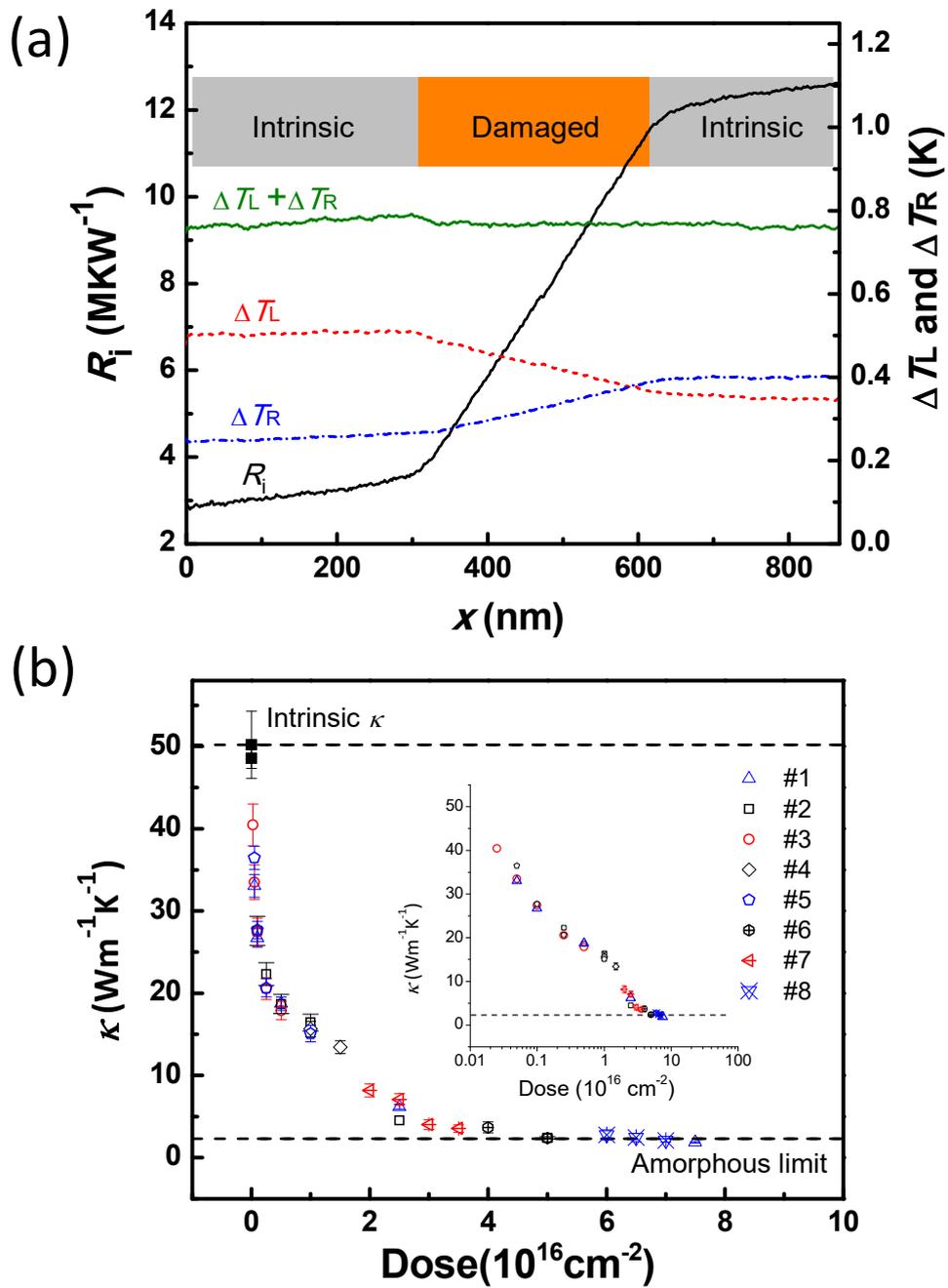

**Figure 2 | Thermal conductivity measurement by e-beam technique. (a)** Measurement result for the damaged portion with highest dose ($7.5\times10^{16}$ cm$^{-2}$) in Sample #1. The solid black line is

the cumulative thermal resistance $R_i(x)$ from the electron beam position $x$ to the left sensor. The dashed red (blue) line is the temperature rise of the left (right) sensor, $\Delta T_L$ ($\Delta T_R$). The solid green line is $\Delta T_L + \Delta T_R$, which is proportional to the power absorption from the travelling electrons. **(b)** Measured thermal conductivity ($\kappa$) of Samples #1 to #8 vs dose. Inset: the same data plotted on a logarithmic scale. The solid black square denotes the thermal conductivity of intrinsic nanowires (namely, with zero dose), that is derived from averaging calculated ones by fitting several intrinsic portions for the eight samples.

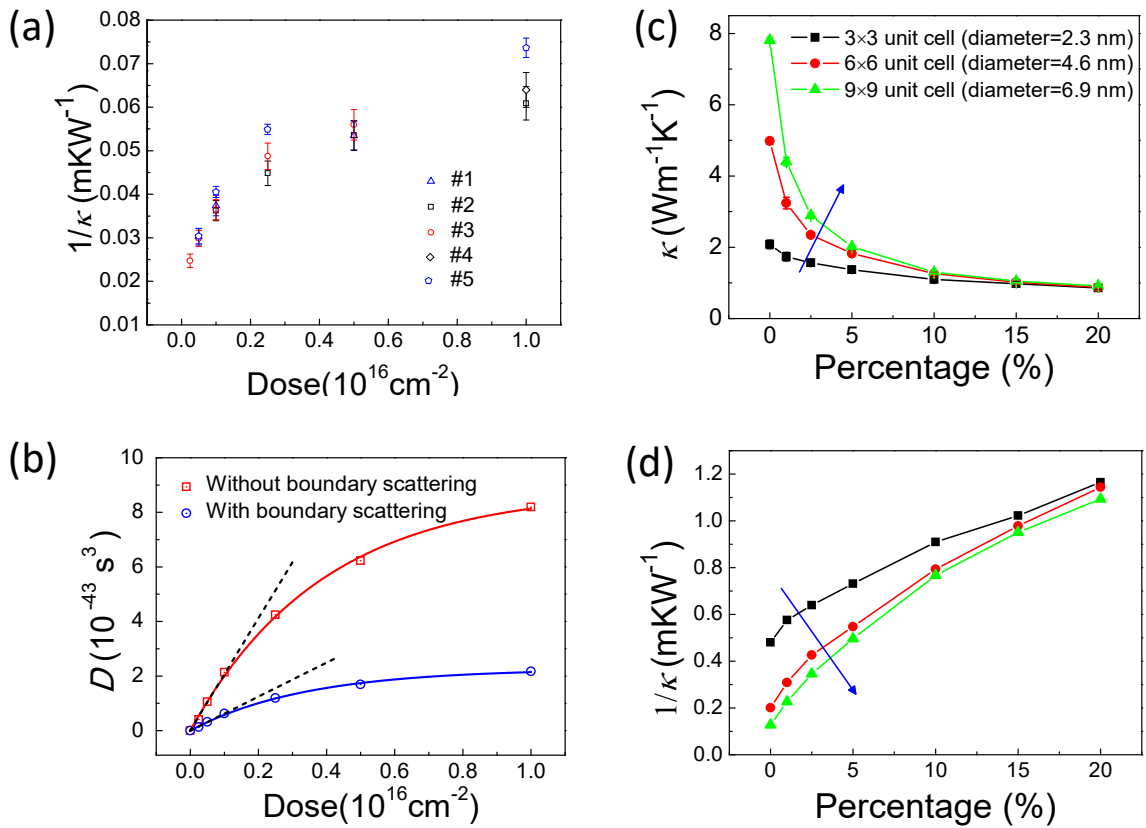

**Figure 3 | Analysis thermal conductivity dependence on dose and vacancy concentration.**
**(a)** Measured thermal resistivity ($1/\kappa$) as a function of dose for low dose regime. For dose <

$1\times10^{15}$ cm$^{-2}$, thermal resistivity increases linearly with dose; whereas as the dose further increases, the thermal resistivity tends to saturate. **(b)** Extracted parameter $D$ as a function of dose with & without considering boundary scattering. Open blue circles and red squares correspond to optimal $D$ fitted from measured thermal conductivities with and without considering the boundary scattering, respectively. $D$ can be well fitted in to an exponential curve $D = E(1 - e^{-F \cdot dose})$, denoted by the red and blue solid line; while the black dashed line represents the linearly fitted curve of $D$ as a function of dose in the low-dose regime: $D = EF \cdot dose$. **(c)** Thermal conductivity as a function of percentage of removed Si atoms from NEMD simulations. The blue arrow indicates the trend with increasing cross sectional area. **(d)** Thermal resistivity ($1/\kappa$) as a function of percentage of removed Si atoms from NEMD simulations. From the calculation in Supplementary Note 1, dose of $1\times10^{16}$ cm$^{-2}$ corresponds to 4% removed Si atoms. The legend is the same as that in (c).

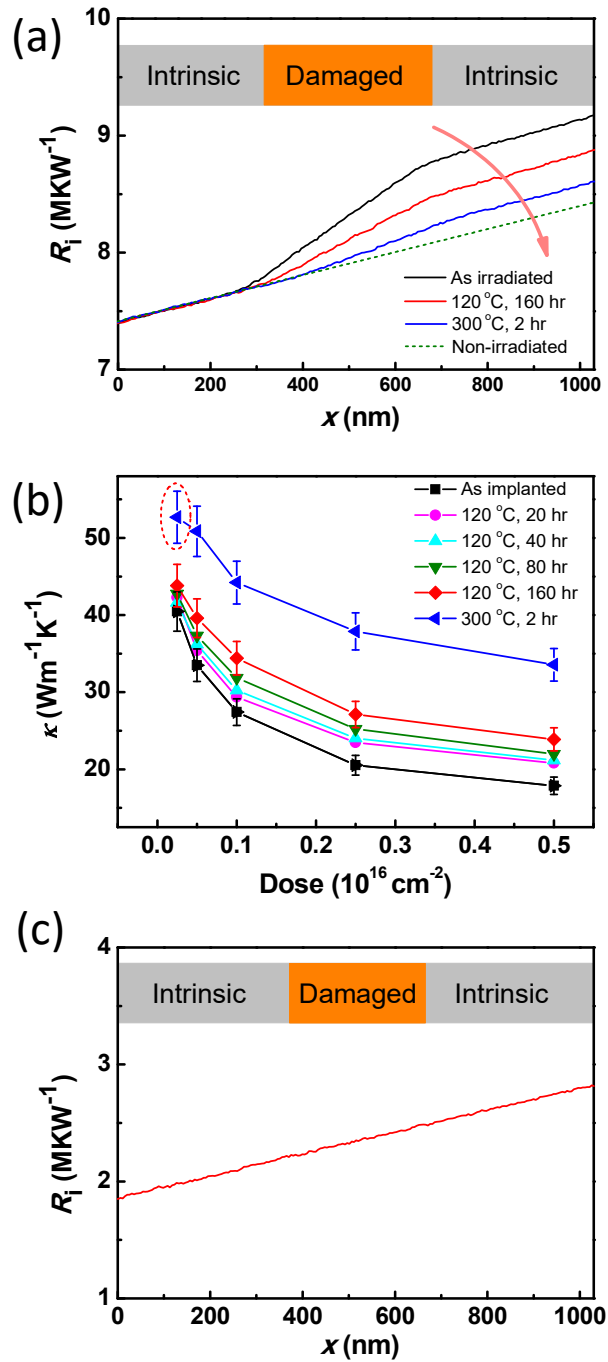

**Figure 4 | Thermal conductivity enhancement due to annealing. (a)** Cumulative thermal resistance ($R_i$) as a function of position, before and after annealing. The damaged portion is

irradiated by helium ions at a dose of $5\times10^{15}$ cm$^{-2}$. The dashed $R_i$ curve illustrates that the portion is not irradiated. **(b)** Thermal conductivity versus dose under different annealing conditions. **(c)** After annealing at 300 °C for two hours, the thermal conductivity of the damaged potion with $2.5\times10^{14}$ cm$^{-2}$ (shown in the red dotted circle in (b)) was recovered.

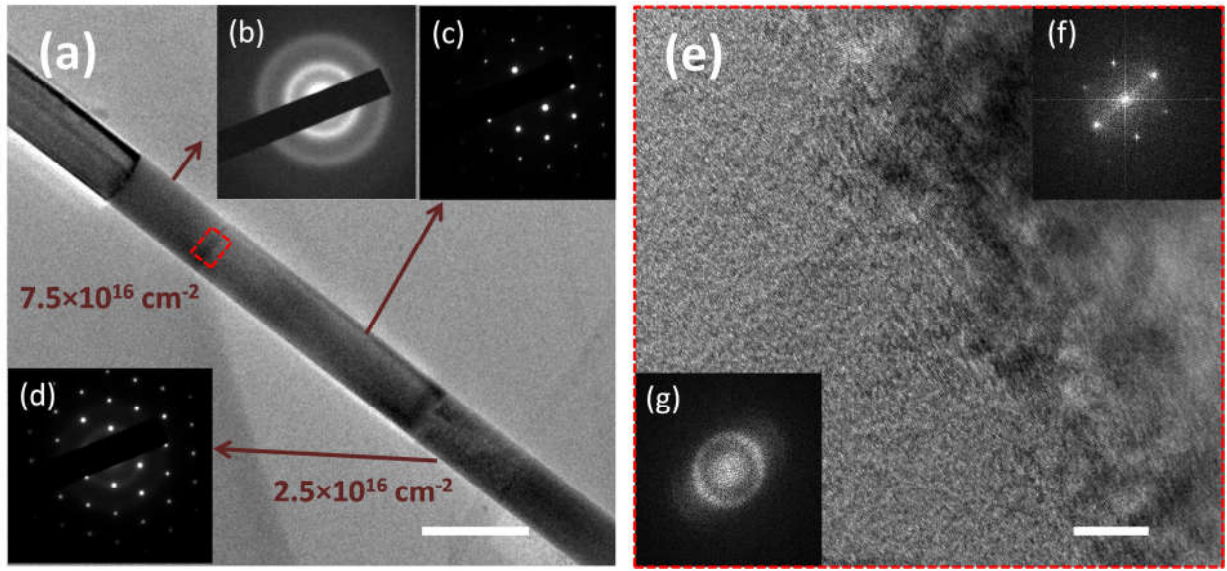

**Figure 5 | TEM characterization.** (a) Bright field TEM image and diffraction patterns of a Si nanowire irradiated by helium ions. Inset (b) is the diffraction pattern of the portion with helium ion dose of $7.5\times10^{16}$ cm$^{-2}$. The Debye-Scherrer rings indicate the fully amorphous status. (d) Diffraction pattern of the portion with helium ion dose of $2.5\times10^{16}$ cm$^{-2}$. Both Debye-Scherrer rings and diffraction spots can be seen, indicative of the coexistence of both amorphous and crystalline phases. (c) Diffraction pattern of non-irradiated portion; **(e)** HRTEM for the selected area marked in red dashed line in (a). Inset (f) and (g) are FFT for crystalline and amorphous regions. The distinguishable interface between irradiated- & non-irradiated areas illustrates the well-controlled dose irradiation. Scale bar for (a) and (e) are 300nm and 10nm, respectively.

AUTHOR INFORMATION

**Corresponding Authors**

* Email: (John T L Thong): elettl@nus.edu.sg

* Email: (Baowen Li): Baowen.Li@colorado.edu

## Acknowledgements

The authors thank Dr. Rongguo Xie for useful discussions. This work is funded by grant MOE2011-T2-1-052 from the Ministry of Education, Singapore, and grant NRF-CRP002-050 from the National Research Foundation, Singapore.

## Author contributions

Y.Z., D.L. and J.T.L.T. designed the experiment; Y.Z. and D.L. made the samples and performed the measurement; Y.Z., D.L., B.L. and J.T.L.T. wrote the paper; J.C. and L.Z. conducted the simulations; Y.Z., A.B., O.S.O., M.J.B., S.K., H.H. and D.S.P. carried out the irradiation works; Y.Z. and R.R.U. performed the TEM characterization; B.L. and J.T.L.T. directed the project.

## Additional information

**Supplementary Information** accompanies this paper at http://www.nature.com/naturecommunications

**Competing financial interests:** The authors declare no competing financial interest.

**Reprints and permission** information is available online at http://npg.nature.com/reprintsandpermissions/